\documentclass[twocolumn,amsmath,amssymb]{revtex4}

\def\gsim{ \lower .75ex \hbox{$\sim$} \llap{\raise .27ex
\hbox{$>$}} }
\def\lsim{ \lower .75ex \hbox{$\sim$} \llap{\raise .27ex
\hbox{$<$}} }

\usepackage{graphicx}
\usepackage{dcolumn}
\usepackage{bm}
\usepackage{graphicx}

\begin{document}

\title{Chameleon Fields: Awaiting Surprises for Tests of
Gravity in Space}

\author{Justin Khoury and Amanda Weltman}

\affiliation{ISCAP, Columbia University, New York, NY 10027,
USA}

\begin{abstract}

We present a novel scenario where a scalar field acquires a
mass which depends on the local matter density: the field is
massive on Earth, where the density is high, but is
essentially free in the solar system, where the density is
low. All existing tests of gravity are satisfied. We predict
that near-future satellite experiments could measure an
effective Newton's constant in space different than that on
Earth, as well as violations of the equivalence principle
stronger than currently allowed by laboratory experiments.

\end{abstract}

\maketitle

Recent observations suggest the existence of a scalar field
which is presently evolving on cosmological time scales. Indeed, the Universe 
is undergoing a period of accelerated expansion as a result of a dark energy
component with negative pressure. Although the current data
is consistent with this being a cosmological constant, the
dark energy is more generally modeled as
quintessence~\cite{quint}: a scalar field rolling down a
flat potential. In order for such a scalar field to evolve cosmologically
today, its mass must be of order $H_0$, the present Hubble
parameter. Thus, one would naively expect it to be
essentially massless on solar system scales, in which case
tests of the Equivalence Principle (EP)~\cite{willbook}
would constrain its coupling to matter to be unnaturally
small.

In this Letter, we propose a novel scenario which allows
scalar fields to evolve on cosmological time scales today
while having couplings of order unity to matter, as expected
from string theory. The idea is that the mass of the scalar
field is not constant in space and time, but rather depends
on the environment, in particular on the local matter
density. Thus, in regions of high density, such as on Earth,
the mass of the field can be sufficiently large to satisfy
constraints on EP violations and fifth force. Meanwhile, on
cosmological scales where the matter density is $10^{30}$
times smaller, the mass of the field can be of order $H_0$,
thus allowing the field to evolve cosmologically today. The
philosophy, therefore, is that cosmological scalar fields,
such as quintessence, have not yet been detected in local
tests of the EP because we happen to live in a dense
environment. Since their physical characteristics depend
sensitively on their environment, we dub such scalar fields:
chameleons. 

In our model, the strength of EP violations and the
magnitude of the fifth force mediated by the chameleon can
be drastically different in space than in the laboratory. In
particular, we find exciting new predictions for near-future
satellite experiments, such as SEE,
$\mu$SCOPE, GG and STEP~\cite{STEP},
that will test gravity in space. We find that it is possible
for SEE to measure an effective Newton's constant that
differs by order unity from the value measured on Earth.
Moreover, the $\mu$SCOPE, GG and STEP satellites could
detect violations of the EP larger than currently allowed by
laboratory experiments. 
Such outcomes would strongly suggest
that a chameleon-like model is realized in Nature. 
Furthermore, they strengthen the scientific case for these missions.
Some of the results below require lengthy calculations presented in a companion paper~\cite{long}. The cosmology is studied elsewhere~\cite{cosmo}.

\begin{figure}
\includegraphics[width=2in]{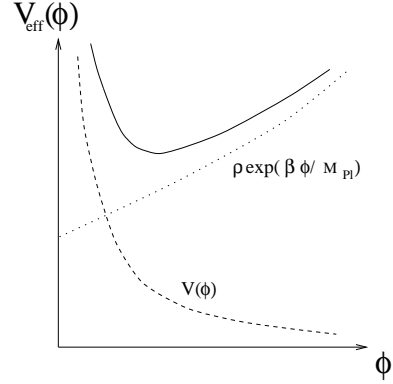}
\caption{The chameleon effective potential $V_{eff}$ (solid
curve) is the sum of a scalar potential $V(\phi)$ (dashed
curve) and a density-dependent term (dotted curve).}
\label{poteff}
\end{figure}

Consider the general lagrangian
\begin{equation}
{\cal L} = \sqrt{-g}\left\{-\frac{M_{Pl}^2{\cal
R}}{2} +\frac{(\partial\phi)^2}{2}+ V(\phi)\right\} 
+ {\cal L}_m(\psi^{(i)},g_{\mu\nu}^{(i)})\,,
\label{action}
\end{equation}
where $M_{Pl}\equiv (8\pi G)^{-1/2}$ is the reduced Planck
mass. Each matter field $\psi^{(i)}$, labeled by $i$,
couples to a metric $g_{\mu\nu}^{(i)}$ related to the
Einstein-frame metric $g_{\mu\nu}$ by a conformal
transformation:
$g_{\mu\nu}^{(i)}=\exp(2\beta_i\phi/M_{Pl})g_{\mu\nu}$,
where $\beta_i$ are dimensionless constants. In harmony with
string theory, we allow the $\beta_i$'s
to be of order unity and to assume different values for
different matter species. 

The potential $V(\phi)$ is assumed to be of the runaway
form. That is, it is monotonically decreasing and
satisfies:
$V,\;V_{,\phi}/V,\;V_{,\phi\phi}/V_{,\phi}\ldots\rightarrow
0$ as $\phi\rightarrow\infty$, as well as
$V,\;V_{,\phi}/V,\;V_{,\phi\phi}/V_{,\phi}\ldots\rightarrow
\infty$ as $\phi\rightarrow 0$. See the dashed curve in
Fig.~\ref{poteff}. A prototypical example is the inverse
power-law potential: $V(\phi) = M^{4+n}\phi^{-n}$, where $n$
is positive and $M$ has units of mass. The runaway form is
generic to non-perturbative potentials in string theory and
is also desirable for quintessence models~\cite{zlatev}.

For simplicity, we focus on a single matter component coupled to a metric $\tilde{g}_{\mu\nu}=\exp(2\beta\phi/M_{Pl})g_{\mu\nu}$.
For non-relativistic matter, one has $\tilde{g}^{\mu\nu}T_{\mu\nu}\approx -\tilde{\rho}$,
where $T_{\mu\nu} = (2/\sqrt{-\tilde{g}})\delta {\cal L}_m/\delta \tilde{g}^{\mu\nu}$ and $\tilde{\rho}$ are the stress tensor and corresponding energy density, respectively.
For convenience, however, we express our equations in terms of $\rho\equiv \tilde{\rho}e^{3\beta\phi/M_{Pl}}$ which is conserved in Einstein frame and hence independent of $\phi$.
Equation~(\ref{action}) then gives
\begin{equation}
\nabla^2\phi = V_{,\phi} +
\frac{\beta}{M_{Pl}}\rho e^{\beta\phi/M_{Pl}}\,.
\label{eom}
\end{equation}

The key realization from Eq.~(\ref{eom}) is that the dynamics of
the chameleon are not governed by $V(\phi)$, but rather by
an {\it effective} potential
\begin{equation}
V_{eff}(\phi) \equiv V(\phi) + 
\rho e^{\beta\phi/M_{Pl}}\,,
\label{veff}
\end{equation}
which is an explicit function of $\rho_i$. Moreover,
although $V(\phi)$ is monotonically decreasing, $V_{eff}$
has a minimum if $\beta>0$. This is shown in
Fig.~\ref{poteff}. In particular, the value of $\phi$ at the
minimum, $\phi_{min}$, and the mass of small fluctuations
about the minimum, $m_{min}$, both depend on $\rho$. More
precisely, $\phi_{min}$ and $m_{min}$ are decreasing and
increasing functions of $\rho$, respectively. That is, the
larger the density of the environment, the larger the mass
of the chameleon.

{\it Solution for a compact object.} We
derive an approximate solution for $\phi$ for a compact object.
For simplicity, we restrict our analysis to the static case and consider a
spherically-symmetric body of radius $R_c$, homogeneous
density $\rho_c$ and total mass $M_c = 4\pi\rho_cR_c^3/3$.
Ignoring the backreaction on the metric, Eq.~(\ref{eom})
reduces to
\begin{equation}
\frac{d^2\phi}{dr^2} + \frac{2}{r}\frac{d\phi}{dr} =
V_{,\phi} +
\frac{\beta}{M_{Pl}}\rho (r) e^{\beta\phi/M_{Pl}}\,.
\label{sun1}
\end{equation}
The density, $\rho(r)$, is equal to $\rho_c$ for $r<R_c$ and
to $\rho_\infty$ for $r>R_c$, where $\rho_\infty$ denotes
the surrounding homogeneous matter density. 

We denote by $\phi_c$ and $\phi_\infty$ the value of $\phi$
that minimizes $V_{eff}$ with $\rho=\rho_c$ and
$\rho_\infty$, respectively. The respective masses of small
fluctuations are $m_c$ and $m_\infty$. The boundary
conditions specify that the solution be non-singular at the
origin ($d\phi/dr=0$ at $r=0$), and that the force on a test
particle vanishes at infinity ($\phi\rightarrow \phi_\infty$
as $r\rightarrow \infty$).

\begin{figure}
\includegraphics[width=2in]{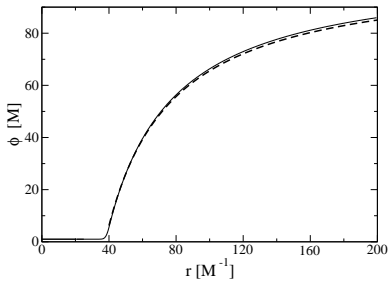}
\caption{Example of solution with thin shell.}
\label{thinshellyes}
\end{figure}

\begin{figure}
\includegraphics[width=2in]{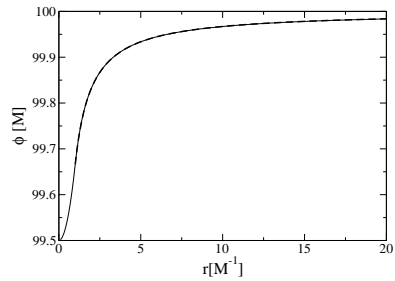}
\caption{Example of solution without thin shell.}
\label{thinshellno}
\end{figure}

For sufficiently large objects, the solution can be
described as follows. Within the object,
$r<R_c$, the field minimizes $V_{eff}$, and thus
$\phi\approx\phi_c$. This holds true everywhere inside the
object except within a thin 
shell of thickness $\Delta R_c$ below the surface where the
field grows. Outside the object, $r>R_c$, the profile for
$\phi$ is essentially that of a massive scalar, $\phi \sim
\exp(-m_{\infty}r)/r$, and tends to $\phi_\infty$ for $r\gg
R_c$, as required by the second boundary condition.

A detailed calculation~\cite{long} shows that the thickness
of the thin shell, $\Delta R_c$, is related to $\phi_{\infty}$, $\phi_c$
and the Newtonian potential of the object, $\Phi_c=M_c/8\pi
M_{Pl}^2R_c$, by
\begin{equation}
\frac{\Delta R_c}{R_c} \approx \frac{\phi_\infty-\phi_c}{6\beta
M_{Pl}\Phi_c}\,.
\label{DR}
\end{equation}
Moreover, the exterior solution ($r>R_c$) is given
by~\cite{long}
\begin{equation}
\phi(r)\approx -\phi_\infty\left(1-\frac{\phi_\infty-\phi_c}{6\beta
M_{Pl}\Phi_c}\right)\frac{R_ce^{-m_\infty (r-R_c)}}{r} + \phi_\infty\,.
\label{interstep}
\end{equation}
Assuming that the density contrast is high, $\phi_c\ll\phi_\infty$, and in the limit that the shell is thin, $\Delta R_c/R_c\ll 1$, Eqs.~(\ref{DR}) and~(\ref{interstep}) combine to give, for $r>R_c$,
\begin{equation}
\phi(r)\approx -\left(\frac{\beta}{4\pi
M_{Pl}}\right)\left(\frac{3\Delta
R_c}{R_c}\right)\frac{M_ce^{-m_\infty (r-R_c)}}{r} +
\phi_\infty\,.
\label{thinsoln}
\end{equation}

The above only applies to objects satisfying the thin-shell condition: $\Delta R_c/R_c\ll 1$. From Eq.~(\ref{DR}), whether or not this condition is satisfied depends
on the ratio of the difference in $\phi$ potential, $\phi_\infty-\phi_c$, to the Newtonian potential of the object, $\Phi_c$. In particular, for fixed $\phi_\infty-\phi_c$ ({\it i.e.}, fixed density contrast), then the more massive the object, the easier it is to satisfy this condition.

Objects with $\Delta R_c/R_c \;\gsim\;1$, however, do not satisfy the thin-shell condition. 
Instead, one has $\phi\sim \phi_\infty$ everywhere in this case, and the exterior solution is
\begin{equation}
\phi(r)\approx -\left(\frac{\beta}{4\pi
M_{Pl}}\right)\frac{M_ce^{-m_\infty (r-R_c)}}{r} + \phi_\infty\,.
\label{warmup}
\end{equation}
Comparison with Eq.~(\ref{thinsoln}) shows that the $\phi$-profile outside large objects is suppressed by a factor of $\Delta R_c/R_c\ll 1$.

The thin shell effect is a consequence of the non-linearity of Eq.~(\ref{sun1}). It follows from requiring the above boundary conditions as well as continuity of $\phi$ and $d\phi/dr$ at $r=R_c$. Satisfying these conditions for sufficiently large objects inevitably leads to a thin shell.

This is confirmed by numerical calculations. Consider, {\it e.g.}, $V(\phi) = M^{5}/\phi$ with $M\approx 6\;{\rm mm}^{-1}$ and $\beta=1$. (We will find that $M\;\lsim \; 1$~mm$^{-1}$, so this is a realistic choice.) The densities are $\rho_c=1\;{\rm g}/{\rm cm}^3$ and $\rho_\infty=10^{-4}{\rm g}/{\rm cm}^3$, mimicking a ball of Be in air, corresponding to $\phi_c/M=1$ and $\phi_\infty/M=100$, respectively. Figure~\ref{thinshellyes} shows the numerical solution for an object of radius $R_c=40\;M^{-1}$. Eq.~(\ref{DR}) predicts a thin shell of thickness $\Delta R_c/R_c \approx 0.0625$ in this case. This is confirmed by the numerics. Indeed, we see from Fig.~\ref{thinshellyes} that $\phi\approx \phi_c$ everywhere inside the object, except within $38\;\lsim\;rM\;\lsim \; 40$. Moreover, the dotted line is a plot of Eq.~(\ref{interstep}) and agrees to within 2\% with the numerics.

Figure~\ref{thinshellno} is the solution for an object with $R_c=M^{-1}$. The numerics confirm that there is no thin shell in this case. Indeed, $\phi\sim \phi_\infty$ everywhere. Moreover, the dotted line is a plot of Eq.~(\ref{warmup}) and is barely distinguishable from the numerical solution. This proves unambiguously that the thin shell effect is real, and that the above expressions provide very good approximations to the actual solution. We should stress that these conclusions are not specific to the particular values of $n$, $\beta$ and $M$ chosen here.

Let us apply these results to the Earth, crudely modeled as a sphere of radius
$R_\oplus$ and density $\rho_{\oplus}=10\;{\rm g}/{\rm
cm}^3$, with an atmosphere $10$ km thick with $\rho_{atm}\approx  10^{-3}\;{\rm g}/{\rm cm}^3$. 
Far away the matter density is approximately that of
baryonic gas and dark matter in our neighborhood of the
Milky Way: $\rho_G\approx 10^{-24}\;{\rm g}/{\rm cm}^3$.
Henceforth, $\phi_\oplus$, $\phi_{atm}$ and $\phi_G$ denote
the value of $\phi$ which minimizes $V_{eff}$ for
$\rho=\rho_\oplus$, $\rho_{atm}$ and $\rho_G$, respectively.
The respective masses are $m_\oplus$, $m_{atm}$ and $m_G$.

The Earth must have a thin shell, for otherwise unacceptably large violations of the EP
will ensue~\cite{long}. Thus the $\phi$-field outside the
Earth is given by Eq.~(\ref{thinsoln}) with
$\phi_\infty=\phi_G$ and $m_\infty=m_G$:
\begin{equation}
\phi(r)\approx -\left(\frac{\beta}{4\pi M_{Pl}}\right)
\left(\frac{3\Delta R_\oplus}{R_\oplus}\right)
\frac{M_\oplus e^{-m_G r}}{r} + \phi_G\,,
\label{earth}
\end{equation}
where $\Delta R_\oplus/R_\oplus =
(\phi_G-\phi_\oplus)/6\beta M_{Pl}\Phi_{\oplus}\ll 1$. In
fact, not only must the Earth have a thin shell, but so must
the atmosphere. This results in a more stringent
condition~\cite{long}
\begin{equation}
\frac{\Delta R_\oplus}{R_\oplus} =
\frac{\phi_G-\phi_\oplus}{6\beta M_{Pl}\Phi_\oplus} <
10^{-7}\,.
\label{condatm}
\end{equation}
It then follows that $\phi\approx\phi_{atm}$ in the
atmosphere. 

For an inverse power-law potential,
$V(\phi)=M^{4+n}\phi^{-n}$, Eq.~(\ref{condatm}) can be
translated into a constraint on the scale $M$ which, for $n$
and $\beta$ of order unity, is given by~\cite{long}:
\begin{equation}
M\;\lsim \; 10^{-3}\; {\rm eV}\approx(1\;{\rm mm})^{-1}\,.
\label{condM}
\end{equation}
Remarkably, this coincides with the energy scale associated
with the dark energy causing cosmic acceleration~\cite{cosmo}. 

Equation~(\ref{condM}) can also be expressed as a bound on
the chameleon interaction range in the atmosphere
($m_{atm}^{-1}$), in the solar system ($m_G^{-1}$) and on
cosmological scales today ($m_0^{-1}$). For $n\;\lsim\; 2$
and $\beta$ of order unity, we find
\begin{eqnarray}
\nonumber
& & m_{atm}^{-1} \;\lsim\; 1\;{\rm mm}-1\;{\rm cm} \\
\nonumber
& & m_G^{-1} \;\lsim\; 10-10^4\;{\rm AU} \\
& & m_0^{-1} \;\lsim\;  0.1-10^3\;{\rm pc}\,.
\label{masses}
\end{eqnarray}
While $\phi$-mediated interactions are short-range in
the atmosphere, $\phi$ is essentially free on
solar system scales. 

{\it Laboratory tests.} We now argue that
Eq.~(\ref{condM}) ensures that laboratory tests of gravity are satisfied. 
Since these are usually performed in vacuum, we need an
approximate solution for $\phi$ inside a vacuum chamber, which we model as a spherical cavity of radius $R_{vac}$. As a boundary condition, we impose $\phi\rightarrow\phi_{atm}$ far from the chamber.
(Note that we use $\phi_{atm}$ rather than $\phi_{\oplus}$ in the laboratory since the field
goes from $\phi_{\oplus}$ to $\phi_{atm}$ within $m_{atm}^{-1}\sim 1\;{\rm mm}$ from the Earth's surface. See Eq.~(\ref{masses}).) Numerical calculations then reveal that, inside the chamber, one has $\phi\approx\phi_{vac}$, where $\phi_{vac}$ satisfies $m^{-1}(\phi_{vac}) \equiv V_{,\phi\phi}^{-1/2}(\phi_{vac}) = R_{vac}$. That is, $\phi_{vac}$ is the field value about which the interaction range is of order $R_{vac}$. Intuitively, this is because $\rho\approx 0$ inside the chamber, and thus the only scale is $R_{vac}$, the size of the chamber.

Hence $\phi$ is essentially free within the vacuum chamber
and generates a fifth-force correction to Newton's constant. If the two test masses used to measure
$G$ have no thin shell, then the correction will be of order
unity for $\beta\sim {\cal O}(1)$, which is clearly ruled
out. Thus the test masses must have a thin shell. If this is
the case, then the $\phi$-field they generate is given by
Eq.~(\ref{thinsoln}) with $\phi_\infty=\phi_{vac}$ and
$m_\infty=R_{vac}^{-1}$, and the correction to $G$ is of
order $(\Delta R_c/R_c)^2$, where $R_c$ is the radius of the
test mass. Therefore, given the current accuracy of
$10^{-3}$ on the value of $G$~\cite{willbook}, this requires 
\begin{equation}
\frac{\Delta R_c}{R_c} = \frac{\phi_{vac}-\phi_c}{6\beta
M_{Pl}\Phi_c} \;\lsim\; 10^{-3/2}\,.
\end{equation}
This ensures that bounds from laboratory searches of a fifth
force and EP violations are satisfied.
For $V=M^{4+n}/\phi^n$ with $\beta$ and $n$ of
order unity, and for typical test bodies inside a vacuum chamber of
$R_{vac}\;\lsim \; 1$ m, it is easy to show that this
condition follows from Eq.~(\ref{condM}). 

{\it Solar system tests.} Tests of GR from solar system data are easily 
satisfied in our model because of the
thin-shell effect which suppresses the $\phi$-force between
large objects. To see this, consider the profile generated
by the Earth given in Eq.~(\ref{earth}). Comparing with
Eq.~(\ref{warmup}), we see that it can be thought of as the
profile for a nearly massless scalar field with effective
coupling
\begin{equation}
\beta_{eff} = 3\beta\cdot \frac{\Delta R_\oplus}{R_\oplus}<
3\beta\cdot 10^{-7}\,,
\label{betaeff}
\end{equation}
where we have used Eq.~(\ref{condatm}) in the last step.
Hence, to describe the effects on planetary motion, we may
think of the chameleon as a free scalar field with coupling
$\beta_{eff}$. Since $\beta_{eff}$ is so small, however, all
bounds from solar system tests of gravity are easily
satisfied~\cite{long}. For instance, consider
light-deflection measurements~\cite{willbook}. 
Treating $\phi$ as a Brans-Dicke field
with effective Brans-Dicke parameter given by $3 +
2\omega_{BD} = (2\beta_{eff}^{2})^{-1}$, we see from
Eq.~(\ref{betaeff}) that the constraint from light-deflection, $\omega_{BD}> 3500$, is
trivially satisfied. 
Similar arguments show that all current constraints from tests of gravity are satisfied~\cite{long}.

{\it Predictions for near-future tests of gravity in space.}
Although $\phi$ mediates short-range interactions on Earth,
we have seen that it is essentially a free field in the
solar system. Thus the magnitude of EP violations and fifth
force in our model are drastically different in space than
on Earth. This opens the door to new and unexpected outcomes
for near-future satellite experiments that will test the EP
and search for a fifth force in space~\cite{STEP}. 

Consider the SEE Project which will measure $G$
in space by accurately determining the orbit of two test
masses at an altitude of $\approx 1000$ km. For a wide range of parameters, we predict that SEE will find a value for $G$ different by ${\cal
O}(1)$ corrections from that measured on Earth, due to fifth
force contributions that are significant in orbit but
exponentially suppressed in the laboratory. 

The {\it sine qua non} for this result is for the
satellite {\it not} to have a thin shell, {\it i.e.},
$\Delta R_{SEE}/R_{SEE} > 1$. The current design for the
capsule has $\Phi_{SEE}\approx 10^{-24}\approx
10^{-15}\Phi_\oplus$. Combining with Eq.~(\ref{condatm}), it
follows that the satellite will fail to have a thin shell if
\begin{equation}
10^{-15} < \frac{\Delta R_\oplus}{R_\oplus} < 10^{-7}\,.
\label{condSEE}
\end{equation}

Equation~(\ref{condSEE}) ensures that the background value
of the chameleon is essentially unperturbed by the satellite
and thus that the $\phi$-mediated force is long-range within
the capsule. Hence, the total force, gravitational plus
chameleon-mediated, between two bodies of mass $M_i$ and
coupling $\beta_i$, $i=1,2$, is
\begin{equation}
|\vec{F}| =
\frac{GM_1M_2}{r^2}\left(1+2\beta_1\beta_2\right)\,.
\label{F12}
\end{equation}
It follows, therefore, that SEE will measure
an effective Newton's constant,
$G_{eff} = G (1 +2\beta_1\beta_2)$, {\it which differs by
order unity from the measured value on Earth}.

Similarly, consider the resulting EP violations. The $\phi$
profile within the capsule (see Eq.~(\ref{earth})) results
in an extra acceleration component $a_\phi$ for a test body
with coupling $\beta$ of: $(a_\phi/a_N) \sim \beta^2\Delta
R_\oplus/R_\oplus$, where $a_N$ is the Newtonian
acceleration. For two bodies of different composition, this
yields a relative difference in free-fall acceleration of
\begin{equation}
\eta \equiv \frac{\Delta a}{a}\approx
10^{-4}\beta^2\frac{\Delta R_\oplus}{R_\oplus} \,, 
\label{etaGG}
\end{equation}
where the numerical factor is appropriate for Be and Nb test
masses~\cite{pol} as used in STEP. Combining with
Eq.~(\ref{condSEE}) yields an allowed range of
\begin{equation}
\beta^2\cdot 10^{-19} < \eta < \beta^2\cdot 10^{-11}\,,
\label{range}
\end{equation}
which overlaps with the sensitivity range of STEP
($\eta\;\gsim\; 10^{-18}$), GG ($\eta\;\gsim\; 10^{-17}$)
and $\mu$SCOPE ($\eta\;\gsim\; 10^{-15}$). Amazingly, $\eta$ 
can be larger than $10^{-13}$, the current bound from
the (ground-based) E$\ddot{{\rm o}}$t-Wash
experiment~\cite{eotwash}. 

If the SEE Project measures a value  for $G$ different than
on Earth, or if the STEP satellite finds an EP-violating
signal stronger than allowed by laboratory experiments, this
will constitute a smoking gun for our model, for it would
otherwise be difficult to reconcile the results in space
with those on Earth.

We thank J.R.~Bond, R.~Brandenberger, P.~Brax, C.~van de Bruck, S.~Carroll, T.~Damour, A.-C.~Davis,
G.~Esposito-Far\`ese, G.~Gibbons, B.~Greene, D.~Kabat, A.~Lukas, J.~Murugan, B.A.~Ovrut, M.~Parikh, S.-J.~Rey, K.~Schaalm, C.L.~Steinhardt, N.~Turok, C.M.~Will, T.~Wiseman,
and especially N.~Kaloper and P.J.~Steinhardt for insightful
discussions.
This work was supported by the CU Academic Quality Fund, the
Ohrstrom Foundation (JK), DOE grant DE-FG02-92ER40699 and
the University of Cape Town (AW).

%\nopagebreak

\end{document}